\documentclass[12pt,a4paper]{article}
\usepackage{amsmath}
\usepackage{amsfonts}
\usepackage{amssymb}
\usepackage{makeidx}
\usepackage[pdftex]{graphicx}

\begin{document}

\title{\textbf{Exergy analysis of magnetic refrigeration}}
\author{Umberto Lucia\\I.T.I.S. A. Volta\\Spalto Marengo 42, 15121 Alessandria, italy} 
\date{}
\maketitle
 
\begin{abstract}
One of the main challenges of the industry today is to face its impact on global warming considering that the greenhouse effect problem is not be solved completely yet. Magnetic refrigeration represents an environment-safe refrigeration technology. The magnetic refrigeration is analysed using the second law analysis and introducing exergy in order to obtain a model for engineering application.
\end{abstract}

\section{Introduction}
One of the main challenges of the industry today is to face its impact on global warming [\ref{muller}]; moreover, the greenhouse effect problem is not be solved completely yet. So, in addition to further developing the vapor compression technology, technical physicists and engineers have begun to explore new refrigeration technology such as magnetic refrigeration. The magnetic cooling technology is based on the use of the magnetocaloric effect (MCE)
applied to various metallic materials and new alloys named magnetocaloric materials (MCM). This application uses reversible temperature change in paramagnetic salts to obtain low
temperatures by adiabatic demagnetization [\ref{muller}]. Magnetic refrigeration is an environment-safe refrigeration technology [\ref{yu}]; indeed, magnetic cooling offers an innovative technical solution that will enable to reduce this global warming impact through two ways: first, it will eliminate refrigerant fluids as CFC, HCFCs and other ammonia, thus it will decrease the direct gas emissions; last, it can use the intrinsic better performance of a magnetocaloric material [\ref{muller}]. The magnetic refrigeration does not have ozone-depleting and greenhouse effects for employing magnetic materials as refrigeration media [\ref{yu}]. 

In 1976, Brown first applied the magnetic refrigeration in a room temperature range [\ref{brown}]. The magnetic refrigeration unit can be compact, for the magnetic entropy density of magnetic material is larger than that of refrigerant gas [\ref{yu}]. Based on the concept of the active magnetic regenerator, Zimm developed a magnetic refrigerator in 1996, which used approximately 3 kg of Gd as working material and generated up to 500-600 W cooling power in a 5 T magnetic field [\ref{zimm}].  

The magnetic field of magnetic refrigeration can be supplied by electromagnet, superconductor or permanent magnet, which have no need for compressors with movable components, large rotational speed, mechanical vibration, noise, bad stability and short longevity. The efficiency of magnetic refrigeration [\ref{lucia}] can be 30-60\% of Carnot cycle [\ref{zimm}], whereas the efficiency of vapor compression refrigeration is only 5-10\% of Carnot cycle [\ref{yu}]. Cooling power of 110 W, temperature span of 33 degrees or more between the cold sink and the hot sink [\ref{muller}]. 

In recent years, magnetic refrigeration on the basis of MCE has been greatly developed in the room temperature range. Whether in the range of room temperature or low temperatures, the magnitude of MCE of magnetic material is the key to cooling capacity [\ref{yu}]. The main research work has been dedicated to new magnetocaloric alloys that can operate in ambient temperature range from $-50^{\circ}\text{C}$ to $+65^{\circ}\text{C}$ [\ref{muller}]. After 2002 different prototypes implementing these MCM and alloys have been designed using permanent magnet systems with high field intensities.

In this paper we wish to introduce a theoretical second law analysis and exergy evaluation in a general Brayton magnetic cycle. To do so in Section 2 we will introduce the relation between entropy and exergy, in Section 3 the second law analysis of the magnetic refrigeration and in Section 4 we will evaluate the quantities introduced for engineergin application with simple approximations.

\section{Entropy generation and exergy}
Classical science emphasized equilibrium and stability, while, recently, it was pointed out the role of fluctuations, instability and evolutionary processes:  irreversible processes are observed everywhere symmetry is broken. In thermodynamics the distinction between reversible and irreversible processes has been introduced by using the concept of entropy so that its formulation is fundamental for understanding thermodynamic aspects of self-organization, evolution of order and life that we see in Nature as it is recently pointed out [\ref{cengel}]. 

The introduction of entropy in classical thermodynamics is related to equilibrium state and reversible transformation. In that context, entropy is a state function depending only on the equilibrium state of the system considered and only entropy differences can be evaluated [\ref{8}]. The introduction of entropy generation comes from the necessity to avoid inequalities [\ref{9}] and use only equation from mathematical point of view. Nothing is really \textit{produced} [\ref{8},\ref{9},\ref{10},\ref{12}]. Indeed, the second law states:
\begin{equation}
\oint\frac{dQ}{T}\leq 0
\end{equation}
defining the total entropy as [\ref{grazzini}]:
\begin{equation}
\label{entropia}
S=\int\bigg(\frac{\delta Q}{T}\bigg)_{rev}=\Delta S_{e}+ S_{g}
\end{equation}
then $S_{g}$, elsewhere called $\Delta S_{irr}$, is considered the generated entropy and it is always $S_{g}\geq 0$ and defined as [\ref{12}]:
\begin{equation}
S_{g}=\int_{\tau_1}^{\tau_2}\dot{S}_{g}d\tau
\end{equation}
with
\begin{equation}
\dot{S}_{g}=\frac{dS}{d\tau}-\sum_{i=1}^{n}\frac{\dot{Q}_{i}}{T_{i}}-\sum_{in}\dot{m}_{in}s_{in}-\sum_{out}\dot{m}_{out}s_{out}
\end{equation}
and $\tau_1$ and $\tau_2$ the initial and final time of the process.

The quantity $\Delta S_{e}$ should be better defined as the entropy variation that will be obtained exchanging reversibly the same \textit{fluxes} throughout the system boundaries. Then entropy is not more then a parameter characterizing the thermodynamic state and the term due to internal irreversibility, $S_{g}$, measures how far the system is from the state that will be attained in a reversible way [\ref{8},\ref{12},\ref{u}].

Entropy is known as the fundamental quantity in the second law thermodynamics, with the following properties [\ref{cengel1}]:
\begin{enumerate}
\item the entropy of a system is a measure of the amount of molecular disorder within the system;
\item a system can only generate, not destroy, entropy;
\item the entropy of a system can be increased or decreased by energy transports across the system boundary;
\item the entropy of the state of a system is a measure of the probability of its occurrence.
\end{enumerate}

The definition and identification of the thermodynamic system is fundamental. The concept of random motion was translated into a notion of order and disorder. Energy transfers or conversions are changes of the state of a system. The natural direction of a change in state of a system is from a state of low probability to one of higher probability: disordered states are more probable than ordered ones. This is the property that changes in all the energy transfers and conversions: the entropy of a state of a system depends on its probability [\ref{cengel}]. 

The non-equilibrium statistical mechanics has been used in order to obtain the statistics for entropy generation in the probability space [\ref{12}]. Indeed, the results obtained underlines that the relevant physical quantities for the stochastic analysis of the irreversibility are the probability of the state in the phase space, the Hamiltonian valued at the end points of trajectory, and the time [\ref{u}]. Very often, all transformations refer to the same time interval and time derivative could be avoided as in classical thermodynamics [\ref{grazzini}].

The exergy of a system is defined as the maximum shaft work that could be done by the composite of the system and a specified reference environment that is assumed to be infinite, in equilibrium, and ultimately to enclose all other systems. Typically, the environment is specified by stating its temperature, pressure and chemical composition. Exergy is not simply a thermodynamic property, but rather it is related to the reference environment [\ref{cengel}]. Exergy is defined as the maximum amount of work which can be produced by a system or a flow of matter or energy as it comes to equilibrium with a reference environment. Exergy is a measure of the potential of the system or flow to cause change, as a consequence of not being completely in stable equilibrium relative to the reference environment. Some properties are, here, summarized as follows, but they can be detailed in Refs. [\ref{cengel},\ref{cengel1}]:
\begin{enumerate}
\item a system in complete equilibrium with its environment does not have any exergy;
\item the more a system deviates from the environment, the more exergy it carries;
\item when the energy loses its quality, it results in exergy destroyed;
\item an engineer designing a system is expected to aim for the highest possible technical efficiency at a minimum cost under the prevailing technical, economic and legal conditions, but also with regard to ethical, ecological and social consequences. Exergy is a concept that makes this work a great deal easier;
\item it is a primary tool in best addressing the impact of energy resource utilization on the environment;
\item it is a suitable technique for furthering the goal of more efficient energy-resource use, for it enables the locations, types, and true magnitudes of wastes and losses to be determined;
\item it is an efficient technique revealing whether or not and by how much it is possible to design more efficient energy systems by reducing the inefficiencies in existing systems;
\item exergy is not subject to a conservation law.
\end{enumerate}

Maximal possible conversion of heat $Q$ to work $L_t$, known as exergy content of heat, depends on the temperature $T$ at which heat is available and the temperature level $T_a$ at which the reject heat can be disposed, that is the temperature of the surrounding. The upper limit for conversion is the Carnot efficiency $1-T_{2}/T_{1}$, where $T_{1}$ and $T_{2}$ are, respectively, the higher and lower temperature of the transformation considered [\ref{cengel1}]. Consequently, exergy exchanged is defined as [\ref{cengel1}]:
\begin{equation}
B=\bigg(1-\frac{T_a}{T}\bigg)Q
\end{equation}

Now, considering the relation (\ref{entropia}), it follows:
\begin{equation}
S=\Delta S_{e}+ S_{g}=\int\bigg(\frac{\delta Q}{T}\bigg)_{rev}=\int\frac{1}{T}\delta\bigg[\bigg(1-\dfrac{T_a}{T}\bigg)^{-1}B\bigg]
\end{equation}

As a consequence of the principle of maximum entropy generation and considering that:
\begin{equation}
\begin{gathered}
\delta S_g =0\\
d(\Delta S_e) = 0
\end{gathered}
\end{equation}
it follows that:
\begin{equation}
\frac{1}{T}\delta\bigg[\bigg(1-\frac{T_a}{T}\bigg)^{-1}B\bigg]=0
\end{equation}
and:
\begin{equation}
\frac{dB}{dT}=-\frac{\dfrac{T_a}{T^2}}{1-\dfrac{T_a}{T}}B
\end{equation}

It follows that:
\begin{enumerate}
\item if $T>T_a$ then $dB/dT <0$ and $dB/d\tau$ represents the maximum power generated during the process
\item if $T<T_a$ then $dB/dT >0$ and $dB/d\tau$ represents the minimum power required during the process
\end{enumerate}
>From this result we can argue that exergy output will not balance the exergy input for real processes in open systems since a part of the exergy input is always destroyed according to the Second Law of Thermodynamics for real processes, as it is pointed out also by the entropy generation maximum principle. Moreover, exergy analysis and entropy generation analysis allow us to determine the most efficient process based on wasting and destroying as little available work as possible from a given input of available work.

\section{Entropy analysis of the magnetic refrigeration}
The magnetocaloric effect, which is intrinsic to all magnetic materials, indicates that the paramagnetic or soft ferromagnetic materials expel heat and their magnetic entropy decreases when the magnetic field is applied isothermally; or otherwise absorb heat and their magnetic entropy increase when the magnetic field is reduced isothermally [\ref{yu}].

Important characteristics of a magnetic material are its total entropy $S$ and the entropy of its magnetic subsystem $S_{M}$ (magnetic entropy). Entropy can be changed by variation of the magnetic field, temperature and other thermodynamic parameters.
The entropy of magnet at constant pressure, $S\left(T,H\right)$
is magnetic field and temperature dependent; it consists of the magnetic
entropy $S_{M}\left(T,H\right)$, both magnetic field and temperature
dependent, the lattice entropy $S_{L}\left(T\right)$ and the electronic
entropy $S_{E}\left(T\right)$, both only temperature dependent [\ref{lucia}]:
\begin{equation}
S\left(T,H\right)=S_{M}\left(T,H\right)+S_{L}\left(T\right)+S_{E}\left(T\right)\label{entropy}
\end{equation}

Magnetic refrigerator completes cooling-refrigeration by magnetic material through magnetic refrigeration cycle. In general a magnetic refrigeration cycle consists of magnetization and demagnetization in which heat is expelled and absorbed respectively, and two other benign middle processes. 
\begin{figure}
\centering 
\label{brayton}
\caption{magnetic Brayton cycle}
\includegraphics[width=0.8 \textwidth]{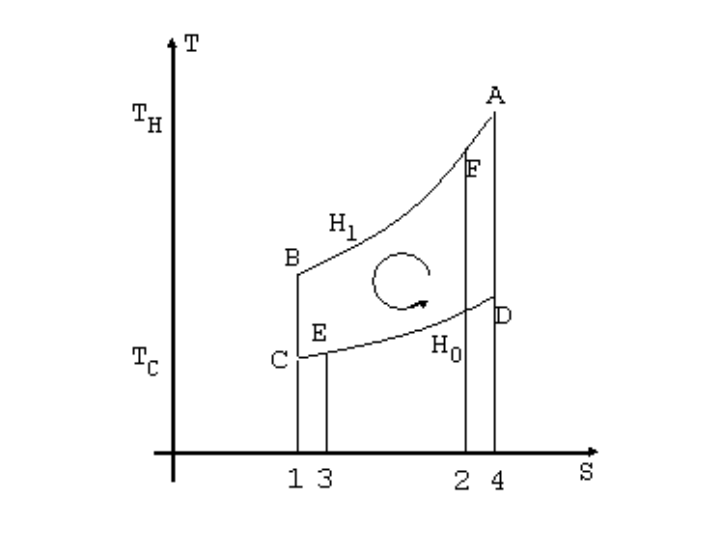}
\end{figure}

In Figure 1 the magnetic Brayton cycle is considered: it consists of two adiabatic processes and two isofield processes. The magnetic refrigerant cycles between the magnetic field of $H_0$ and $H_1$, and the temperature of high and low temperature heat source $T_H$ and $T_C$, respectively. The total entropy and the magnetic part are displayed for an applied external field $H_1$ and $H_0$. During the isofield cooling process $A\rightarrow B$ (constant magnetic field of $H_1$), magnetic refrigerant expels heat of the area of $AB14$. During the isofield heating process $C\rightarrow D$ (constant magnetic field $H_0$), magnetic refrigerant absorbs heat of the area of $DC14$. No heat flows from and out of the magnetic refrigerant during the adiabatic magnetization process $D\rightarrow A$ and the adiabatic demagnetization $B\rightarrow C$ process. The Brayton cycle can exhibit optimal performance as well with magnetic refrigerants having parallel T-S curves [\ref{tishin}]. Since the lattice entropy is too large to neglect in room temperature, part of the refrigeration capacity of the magnetic refrigerant is consumed for cooling the thermal load of lattice system, decreasing the gross cooling capacity of the magnetic refrigerant [\ref{chen}]. By adding a regenerator to the magnetic refrigeration system, the heat expelled by lattice system in one stage of the cycle is restored and returned to lattice system in another stage. So the capacity used for cooling lattice system load can be utilized effectively for the increase of effective entropy change and temperature span.

The magnetic refrigeration ideal $COP$ has recently been studied in [\ref{lucia}] and it is usually defined as:
\begin{equation}
COP=\frac{Q_{0}}{L}
\end{equation}
with $Q_{0}$ the subtracted heat because of the magnetocaloric effect, and $L$ the work done. It is useful to underline that only the magnetic entropy can be controlled by changing
the strength of the magnetic field.

Some considerations have been introduced [\ref{yu},\ref{lucia}]:
\begin{enumerate}
\item magnetization at constant field in both paramagnets and ferromagnets decrease with increasing temperature $(\partial M/\partial T)_H <0$;
\item large total angular momentum number $J$ and Landé factor $g$ of ferromagnetic material, are crucial to magnetocaloric effect;\item modest Debye temperature;\item modest Curie temperature in the vicinity of working temperature to guarantee that the large magnetic entropy change can be obtained in the whole temperature range of the cycle;\item essentially zero magnetic hysteresis; 
\item small specific heat and large thermal conductivity to ensure remarkable temperature change and rapid heat exchange; 
\item	large electric resistance to avoid the eddy current loss; 
\item fine molding and processing behavior to fabricate the magnetic materials satisfactory to the magnetic refrigeration.
\item the $COP$ depends on temperatures in non linear way;
\item it does not depend from the value of the magnetic field, but only
from its variation.
\end{enumerate}

Since the lattice entropy is too large to be neglected in room temperature, part of the refrigeration capacity of the magnetic refrigerant is consumed for cooling the thermal load of lattice system, decreasing the gross cooling capacity of the magnetic refrigerant [\ref{chen}]. By adding a regenerator to the magnetic refrigeration system, the heat expelled by lattice system in one stage of the cycle is restored and returned to lattice system in another stage. So the capacity used for cooling lattice system load can be utilized effectively for the increase of effective entropy change and temperature span.

\section{Exergy analysis of the magnetic refrigeration}
The magnetic Brayton cycle is considered as previously described. During the isofield cooling process $A\rightarrow B$ (constant magnetic field of $H_1$), magnetic refrigerant expels heat $Q_1$ equal to the area of $AB14$ in Figure 1. During the isofield heating process $C\rightarrow D$ (constant magnetic field $H_0$), magnetic refrigerant absorbs heat $Q_0$ equal to the area of $DC14$. No heat flows from and out of the magnetic refrigerant during the adiabatic magnetization process $D\rightarrow A$ and the adiabatic demagnetization $B\rightarrow C$ process. To develop the exergy analysis of this cycle it is necessary to evaluate these areas, but to do so the function $S=S(T,H)$ must be known.

The heat exchanged is:
\begin{equation}
\begin{gathered}
Q_0 = - \int_{B}^{A}TdS = area(DC14)\\
Q_1 = \int_{C}^{D}TdS = area(AB14)
\end{gathered}
\end{equation}

These integrals can be obtained by evaluating in a geometric way the two areas as:
\begin{equation}
\begin{gathered}
Q_0 = area(DC14)= \frac{T_C + T_D}{2}(S_D-S_C)\\
Q_1 = area(AB14)= \frac{T_A + T_B}{2}(S_A-S_B)
\end{gathered}
\end{equation}
where $T_{1m}=(T_A + T_B)/2$ is the mean value of the temperature between $T_A$ and $T_B$ and $T_{0m}=(T_C + T_D)/2$ is the mean value of the temperature between $T_C$ and $T_D$. Considering that $(S_A-S_B)=(S_D-S_C)=\Delta S$ then it follows:
\begin{equation}
\begin{gathered}
Q_0 = T_{0m}\Delta S\\
Q_1 = T_{1m}\Delta S
\end{gathered}
\end{equation}

Consequently, it follows:
\begin{equation}
\begin{gathered}
B_0 = \bigg(1-\frac{T_a}{T_{0m}}\bigg)Q_0= \bigg(1-\frac{T_a}{T_{0m}}\bigg)T_{0m}\Delta S\\
B_1 = \bigg(1-\frac{T_a}{T_{1m}}\bigg)Q_1=\bigg(1-\frac{T_a}{T_{1m}}\bigg)T_{1m}\Delta S\\
L=Q_1-Q_0=T_{1m}\Delta S - T_{0m}\Delta S\\
COP=\frac{Q_0}{L}=\frac{T_{0m}}{T_{1m} - T_{0m}}\\
\eta_{ex}=\frac{B_0}{B_0+B_1}=\frac{T_{0m}-T_a}{T_{0m}+T_{1m}-2T_a}
\end{gathered}
\end{equation}
with $\eta_{ex}$ exergy efficiency.

\section{Conclusions}
In 1881, the study of magnetic refrigeration was started with the
discovery of magnetocaloric effect, when E. Warburg discovered the
thermal effect of metal iron when he applied it in a varying magnetic
field. P. Debye and W.F. Giauque explained the nature of magnetocaloric
effect and suggested an ultra-low temperature can be achieved by adiabatic
demagnetization cooling. Then it has been used in cryogenic
refrigeration since 1930. In 1976, G.V. Brown applied the magnetic
refrigeration in a room temperature range. Magnetic refrigeration
is an environment safe refrigeration technology because it does not
have ozone-depleting and greenhouse effects; in fact, it uses magnetic
materials as refrigeration media. In recent years, magnetic
refrigeration has been greatly developed in the room temperature range
because it is understood that it may be the key of cooling capacity [\ref{lucia}].

The entropy concept and its production in non-equilibrium processes form the basis of modern thermodynamic engineering and technical physics. Entropy has been proved to be a quantity that is related to non-equilibrium dissipative process. The second law of thermodynamics states that for an arbitrary adiabatic process the entropy of the final state is equal to (reversible process) or larger than that of the initial state, what means that the entropy tends to grow because of irreversibility. The MaxEP may be viewed as the natural generalization of the Clausius-Boltzmann-Gibbs formulation of the second law. In the last decades, the fundamental role of entropy generation has been pointed out in the analysis of real systems and an extremum principle for this quantity han been introduced [\ref{8},\ref{12}]. Exergy analysis takes the entropy portion into consideration by including irreversibilities [\ref{cengel}]. 

In this paper a link between entropy generation and exergy has been found. It has been obtained that, as a consequence of the maximum entropy production, $dB/d\tau$ represents the maximum power generated during the process if $T>T_a$, while $dB/d\tau$ represents the minimum power required during the process if $T<T_a$. Moreover the ideal $COP$ and exergy efficiency for the ideal magnetic refrigeration has been obtained by using a linear approach in relation to an ideal general Brayton cycle.

\bibliographystyle{unsrt}


\end{document}